\documentclass[journal]{IEEEtran}
\usepackage{graphicx}
\usepackage{amsmath,amssymb,amsfonts}
\usepackage{xcolor}
\usepackage{csquotes}
\usepackage{textcomp}
\usepackage{gensymb}
\usepackage{hyperref}
\usepackage{multicol}
\usepackage{subcaption}
\usepackage{bm}
\usepackage{setspace}
\usepackage{multirow}
\usepackage{epstopdf}
\usepackage{comment}
\usepackage[font=footnotesize]{caption}
\usepackage{subcaption}
\usepackage[ruled,vlined]{algorithm2e}
\usepackage{stfloats}
\usepackage{steinmetz}
\usepackage{blindtext}
\usepackage{cite}

\begin{document}
\pagestyle{plain}
\title{GridSweep Simulation:  Measuring Subsynchronous Impedance Spectra of Distribution Feeders}
\author{
       Lingling~Fan,~\IEEEmembership{Fellow,~IEEE},
       Zhixin~Miao,~\IEEEmembership{Senior Member,~IEEE},
       Jason~MacDonald,~\IEEEmembership{Member,~IEEE},
       Alex~McEachern,~\IEEEmembership{Life~Fellow,~IEEE}
        \thanks{ This work was performed in part by U.S. Deparment of Energy's (DOE) Lawrence Berkeley National Laboratory, operated by the University of California under Contract No. DE‐AC02‐05CH11231. It was jointly funded by DOE’s Office of Energy Efficiency \& Renewable Energy and its Office of Electricity.  The views expressed in the article do not necessarily represent the views of the DOE or the U.S. Government.
}}
\maketitle
\begin{abstract}
Peaks and troughs in the subsynchronous impedance spectrum of a distribution feeder may be a useful indication of oscillation risk, or more importantly lack of oscillation risk, if inverter-based resource (IBR) deployments are increased on that feeder.  GridSweep is a new instrument for measuring the subsynchronous impedance spectra of distribution feeders.  It combines an active probing device that modulates a 120-volt 1-kW load sinusoidally at a user-selected GPS-phase locked frequency from 1.0 to 40.0 Hz, and with a recorder that takes ultra-high-precision continuous point-on-wave (CPOW) 120-volt synchrowaveforms at 4 kHz. This paper presents a computer simulation of GridSweep's probing and measurement capability. We construct an electromagnetic transient (EMT) simulation of a single-phase distribution feeder equipped with multiple inverter-based resources (IBRs). We include a model of the GridSweep probing device, then demonstrate the model's capability to measure the subsynchronous apparent impedance spectrum of the feeder.  
Peaks in that spectrum align with the system's dominant oscillation modes caused by IBRs. 
\end{abstract}

\IEEEoverridecommandlockouts

\section{Introduction}
\IEEEPARstart{G}{rid}Sweep  
is a new instrument developed for the U.S. Department of Energy
by Alex McEachern at the Lawrence Berkeley National Laboratory.  It measures the subsynchronous impedance spectrum of a distribution feeder or grid.  The instrument conducts active subsynchronous probing at one 120-volt outlet at one location on a distribution feeder, while simultaneously recording GPS-synchronize parts-per-billion voltage synchrowaveforms at another 120-volt outlet which must be at a different location on the same distribution feeder  \cite{mceachern2022gridsweep}.

The GridSweep concept can be traced back to the Chief Joseph Brake, a 1400-MW switchable resistor built in the 1970s for oscillation damping \cite{shelton1975bonneville}. Energizations of the Chief Joseph Brake also provides information about the stability of the WECC system. During 2005 and 2006, real-time phasor measurement unit (PMU) data were captured from the insertion tests and the dominant oscillation modes (N-S mode at 0.2-0.3 Hz, Alberta mode at 0.37 Hz) were identified \cite{hauer2009use}. 

Compared to the Chief Joseph Brake, the GridSweep instrument is tiny, and its 1-kW probing signals result in tiny voltage changes, measured in parts-per-million per-unit or even parts-per-billion per-unit. 

The objective of this paper is to simulate GridSweep's probing and measurement capability in an EMT testbed. We show that GridSweep approach is capable of measuring apparent subsynchronous impedance of a simulated distribution feeder.  
We adopt the terminology ``apparent impedance'' from  \cite{rygg2017apparent}. In \cite{rygg2017apparent}, Rygg and Molinas proposed a method based on injection of a small voltage or current at a location and further defined the apparent impedance as the ratio between the voltage and current at the injection point. Note that this impedance is the equivalent impedance of the entire system viewed from the probing location. In many other scenarios, researchers measure a device or a subsystem's impedance (e.g., \cite{fan2020identifying}), instead of the impedance of an entire system. Apparent impedance indeed reflects the closed-loop system. Therefore, peaks in the apparent impedance spectra implicate oscillation modes and a sharper peak implicates poorly damped oscillations. 

Conventional impedance measurement methods require a probing voltage source or current source. This is the approach taken by the measurement testbed for MW size inverters located in the National Renewable Energy Lab campus  \cite{fan2020identifying}. That testbed requires a controllable voltage source at 13.2 kV. Similarly, the impedance measurement requires current sources in \cite{huang2009small}.  
In contrast, GridSweep has the advantage of not requiring an additional source.  Instead, it can simply be plugged into  120-volt outlets at the edge of the grid, and conduct subsynchronous measurements of feeders and, in some cases, substations. Essentially GridSweep modulates a load. This probing method is similar to those deployed in the lab scale or computer simulation testbeds where transient time-domain data caused by load switching off has been used for impedance identification \cite{valdivia2011impedance, fan2020time}. Compared to the load switching probing method, GridSweep has the advantage of providing controllable and continuous probing signals. 

The non-commercial GridSweep instrument is at technology readiness level (TRL) 9, and has been deployed on real-world distribution grids at several U.S. utilities.  A small fleet of these experimental instruments is maintained at Lawrence Berkeley National Lab for deployment in future experiments.


In the following, we present the theoretic analysis and EMT simulation results to demonstrate GridSweep's capability to capture subsynchronous apparent impedance in the $dq$ frame. While the physical GridSwep device uses a stepped-in-frequency series of single-frequency subsynchronous injections as perturbation, in this simulation we substitute step change injection and continuously-varied chirp signal injection. Prior research \cite{fan2020time,fan2023dq} has shown  parametric models can be quickly identified by use of transient data and advanced system identification algorithms. 
%
%
%
%

\section{Theoretical analysis}

\begin{figure}[!h]
    \centering
    \includegraphics[width=0.9\linewidth]{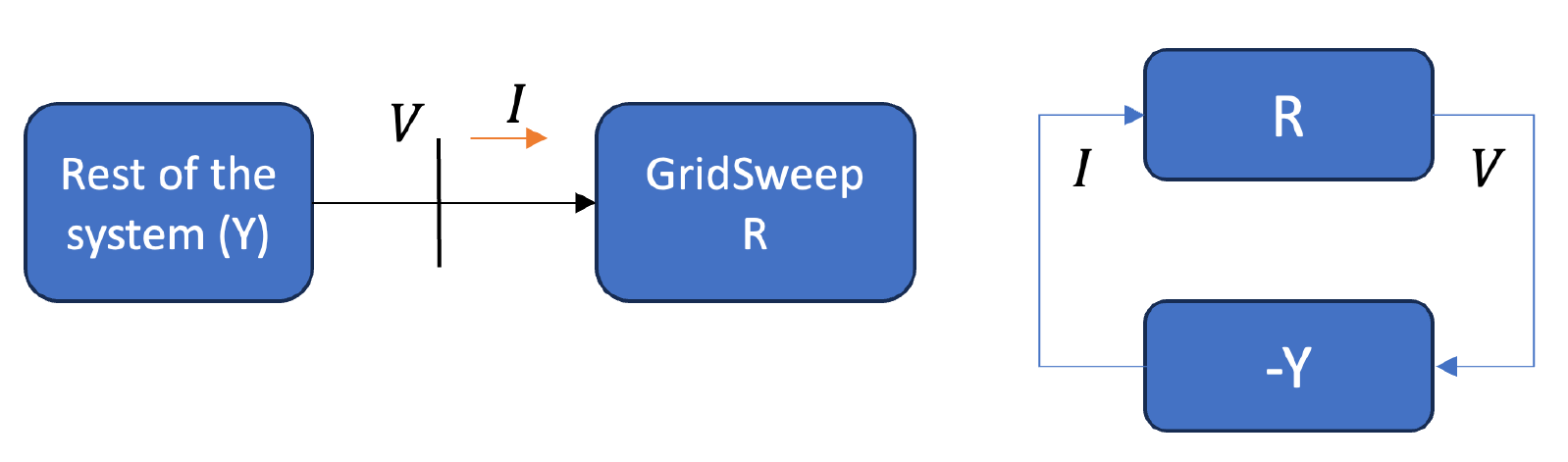}
    \caption{    
    Illustration of the GridSweep, represented by $R$, interacting with the rest of the system, represented by $Y$.}
    \label{fig1}
\end{figure}

GridSweep probing connects a common 1-kW resistive heater, pulse-width modulated at a programmable subsynchronous frequency, located at one 120-volt outlet.  GridSweep recording measures the voltage change at this or another 120-volt outlet. If the probing and measurement locations are the same, the measured voltage change is related to the probing current, as shown in Fig. \ref{fig1}. 

With GridSweep modulating a resistive load at $f_p$ frequency, the resulting continuous point-on-wave (CPOW) voltage measurement contains $(60+f_p)$\textendash Hz and $(60-f_p)$\textendash Hz non-fundamental components due to frequency coupling phenomena \cite{bakhshizadeh2016couplings}, \cite{fan2023modeling} (chapter 3). To directly use the CPOW voltage measurements, we have to deal with two non-fundamental components for a single perturbing subsynchronous signal. Doing so is less convenient than examining voltage's space vector in the $dq$ synchronous frame. If the single-phase voltage measurement is expressed as follows:
\begin{align}
v(t) = \hat{v}(t) \cos( \omega_0 t + \theta(t)),
\end{align}
where $\omega_0$ is the nominal frequency in rad/s. $v(t)$'s  analytical form is notated as 
\begin{align}
\vec{v}(t) = \hat{v}(t) e^{j\theta(t)} e^{j  \omega_0 t }.
\end{align}
 In the $dq$ frame where the $d$-axis is leading the static reference by $\omega_0 t + \theta_0$, the voltage vector $\overline{V}_{dq}$ is expressed as:
\begin{align}
\overline{V}_{dq} = \hat{v} e^{j (\theta- \theta_0)} = \underbrace{\hat{v} \cos(\theta- \theta_0)}_{v_d} +j \underbrace{\hat{v} \sin(\theta- \theta_0)}_{v_q}.
\end{align} 

$v_d$ and $v_q$ can be found directly by averaging the voltage measurement $v(t)$ after multiplication with sinusoidal signals. 
\begin{align}
v_d & = \frac{2}{T}\int^T_0{v(t) \cos(\omega_0 t + \theta_0)}dt = \hat{v} \cos(\theta - \theta_0), \notag \\
v_q & = \frac{2}{T}\int^T_0{v(t) (-\sin(\omega_0 t + \theta_0))}dt = \hat{v} \sin(\theta - \theta_0),
\label{eq:dq}
\end{align}
where $T$ is $2\pi /\omega_0$. 

\begin{figure}[!h]
 \vspace{-0.15in}
    \centering
    \includegraphics[width=0.8\linewidth]{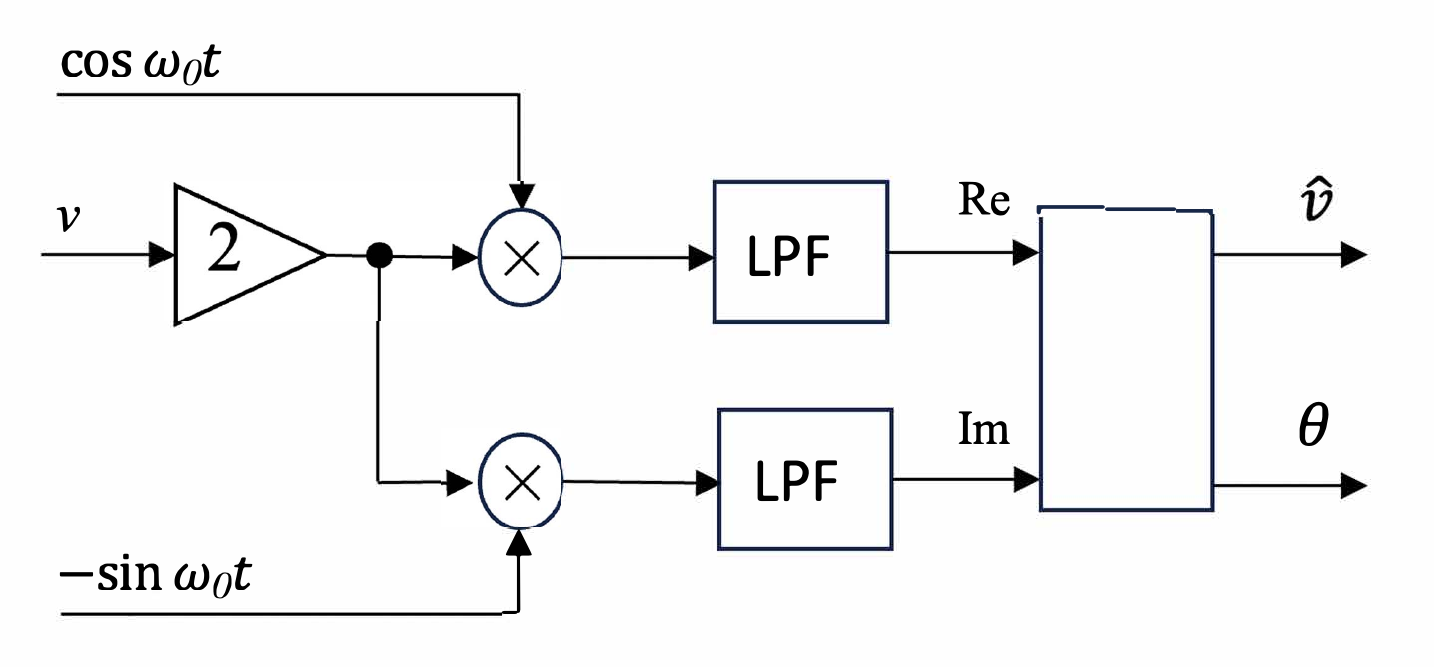}
    \caption{Pre-processing CPOW voltage measurement to extract magnitude and angle in the $dq$ frame.}
    \label{fig_dq}
\end{figure}

Fig. \ref{fig_dq} shows the preprocessing blocks. The voltage CPOW measurement \textit{v} is per unitized and multiplied by 2. Its inner product with the cosine waveform is passed through a low-pass filter (LPF) (e.g., a moving average filter) to have the magnitude associated with $\cos(\omega_0 t)$ computed. Its inner product with the sine waveform is also passed through a low-pass filter to have the coefficient associated with $\sin(\omega_0 t)$ computed. From there, the magnitude in pu and the angle in degree are treated as outputs.

In the $dq$-frame, the voltage and current vector are related as follows:
\begin{align}
 \underbrace{\begin{bmatrix}  v_d \\v_q \end{bmatrix}  }_{\bf V}= \underbrace{\begin{bmatrix}  R & 0 \\ 0 &R \end{bmatrix}  }_{\bf R}\underbrace{\begin{bmatrix}  i_d \\ i_q \end{bmatrix}}_{\bf I} = - \underbrace{\begin{bmatrix}  Z_{dd} & Z_{dq} \\ Z_{qd} & Z_{qq} \end{bmatrix} }_{\bf Z }{\bf I},
\end{align}
 where $\bf Z$ notates the impedance of the rest of the system and its reciprocal is $\bf Y$ (${\bf Y} = {\bf Z}^{-1}$).  $R$ is varying and can be notated as  $R = R_0(1+p)$ where $p$ is the perturbation and $R_0$ is the resistance of the GridSweep before a perturbation is imposed. The linearized relationship between voltage and current in the $dq$-frame becomes as follows:
\begin{align}
\Delta {\bf V} = {R_0} \Delta {\bf I} + {\bf I_0} \Delta { R} = { R_0} \Delta {\bf I} + R_0 {\bf I_0 }  \cdot p,
\end{align}
where subscript $0$ notates the initial condition.

\begin{figure}[!h]
    \centering
       \includegraphics[width=0.7\linewidth]{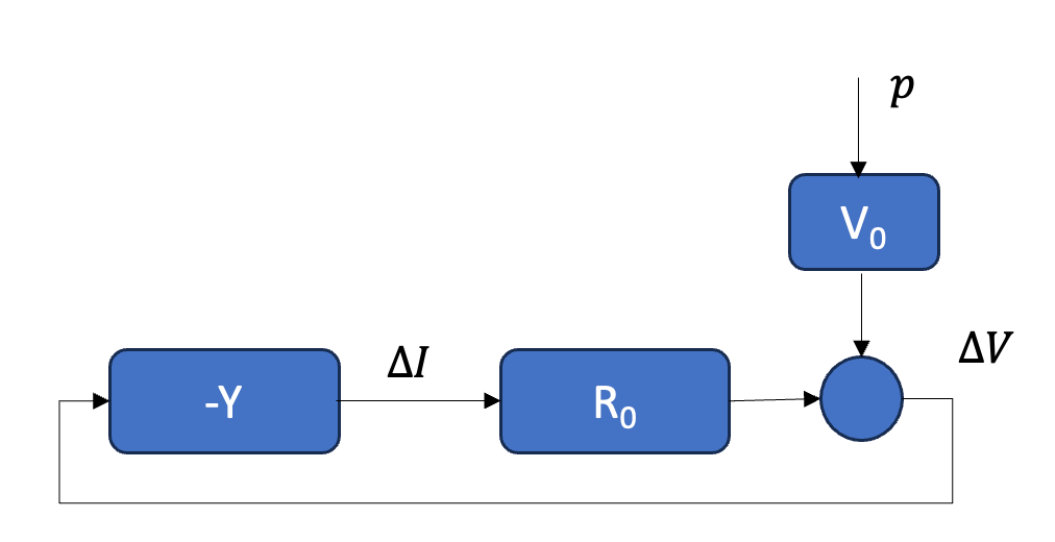}
    \caption{Block diagram of the perturbation and voltage relationship.}
    \label{fig2}
\end{figure}

Note that $R_0 {\bf I_0}$ can be replaced by $\bf V_0$, the initial voltage vector. Therefore, the block diagram describing the relationship between the perturbation and the 
voltage vector is shown as Fig. \ref{fig2}.

Therefore, 
\begin{align}
\frac{\Delta {\bf V}}{p } = \underbrace{({\bf I}+R_0{\bf Y})^{-1}}_{G}{\bf V_0} \approx \frac{1}{R_0} {\bf Z} \cdot {\bf V_0}. 
\label{eq7}
\end{align}

The approximation is made since $R_0$ is very large and all its element of $R_0 {\bf Y}$ is much greater than 1. 
Therefore,
the system is related to the impedance matrix $\bf Z$. If ${\bf V_0}= [V_0, 0]^T$, the input/output models are related to the $dd$ and $qd$ components of $\bf Z$.  

Note that when the $dq$ frame aligns with the voltage vector at the initial steady state, or $\theta= \theta_0$, then according to \eqref{eq:dq},
\begin{align}
\Delta v_d = \Delta \hat{v}, \>\> \Delta v_q = V_0 \Delta \theta 
\end{align}
Therefore, the magnitude and the angle of the voltage are indeed equal or proportional to $v_d$ and $v_q$ viewed in a $dq$-frame aligned with the steady-state voltage vector. 

In short, 
\begin{align}
\begin{bmatrix}  {\Delta \hat{v}}/{p} \\ {\Delta \theta}/{p} \end{bmatrix} = 
 \begin{bmatrix} G_{11}V_0 \\ G_{21} \end{bmatrix} \approx 
 \frac{1}{R_0}
  \begin{bmatrix} Z_{dd} V_0 \\ Z_{qd} \end{bmatrix} 
 \end{align}
 where $G$ is defined in \eqref{eq7} and $G_{11}$ and $G_{21}$ are the elements of the 2-by-2 matrix $G$. 
 
 The above theoretical analysis shows that the magnitude and angle of the fundamental component are related to the two components of the apparent impedance $\bf Z$. 
 
%


\section{Comparison of GridSweep results with Ground Truth}
In Section III and Section IV, we present two sets of simulation results to demonstrate the capability of the GridSweep measurements, respectively. For the first set testbeds presented in Section III, we can find the ground truth since the results can be analytically determined. The GridSweep measurements will be compared with the ground truth. 
For the second set presented in Section IV, we do not have knowledge on ground truth and will resort to other measures for comparison.

In this section, we perform simulations using RLC source impedances, because we know that the results can be analytically determined.  We investigate three cases of RLC source impedances: (1) an ideal AC voltage source followed by a simple RLC source impedance, probed by the GridSweep-simulating variable resistor, as shown in Fig. \ref{fig:testbed1}; (2) we move the source impedance capacitor to a shunt configuration, as shown in Fig. \ref{fig:testbed2}; and (3) we explore multiple simulated GridSweep CPOW recordings at locations other than the probing location, as shown in Fig. \ref{fig:testbed3}.  In each of these three cases, we compare the simulated measured results to the analytically calculated ground truth.   

We use the three cases to demonstrate the following:
\begin{itemize}
\item GridSweep measurements reflect the apparent impedance components combined with the effect of a low-pass filter in the signal processing. 
\item Peaks in the GridSweep measurement spectra correspond to oscillation modes. 
\item If the probing location and measurement location are different, GridSweep's measurements reflect ratio of the components of the apparent impedance
\end{itemize}

\subsection{Case 1: Measure an RLC circuit}

In case 1, the impedance of the RLC circuit in the static frame is $R+Ls + 1/(Cs)$. In the $dq$ frame, the impedance matrix $\bf Z$ (reflecting the relationship between the voltage measurement $v$ and the current injection $i$) is expressed as follows:
\begin{align}
{\bf Z} = \begin{bmatrix}  R+ Ls+\frac{s}{C(s^2+\omega_0^2)} & -\omega_0 L + \frac{\omega_0}{C(s^2+\omega_0^2)} \\
\omega_0 L - \frac{\omega_0}{C(s^2+\omega_0^2)} & R+ Ls+\frac{s}{C(s^2+\omega_0^2)}
\end{bmatrix} 
\end{align}
\begin{figure}[!h]
    \centering
        \includegraphics[width=0.7\linewidth]{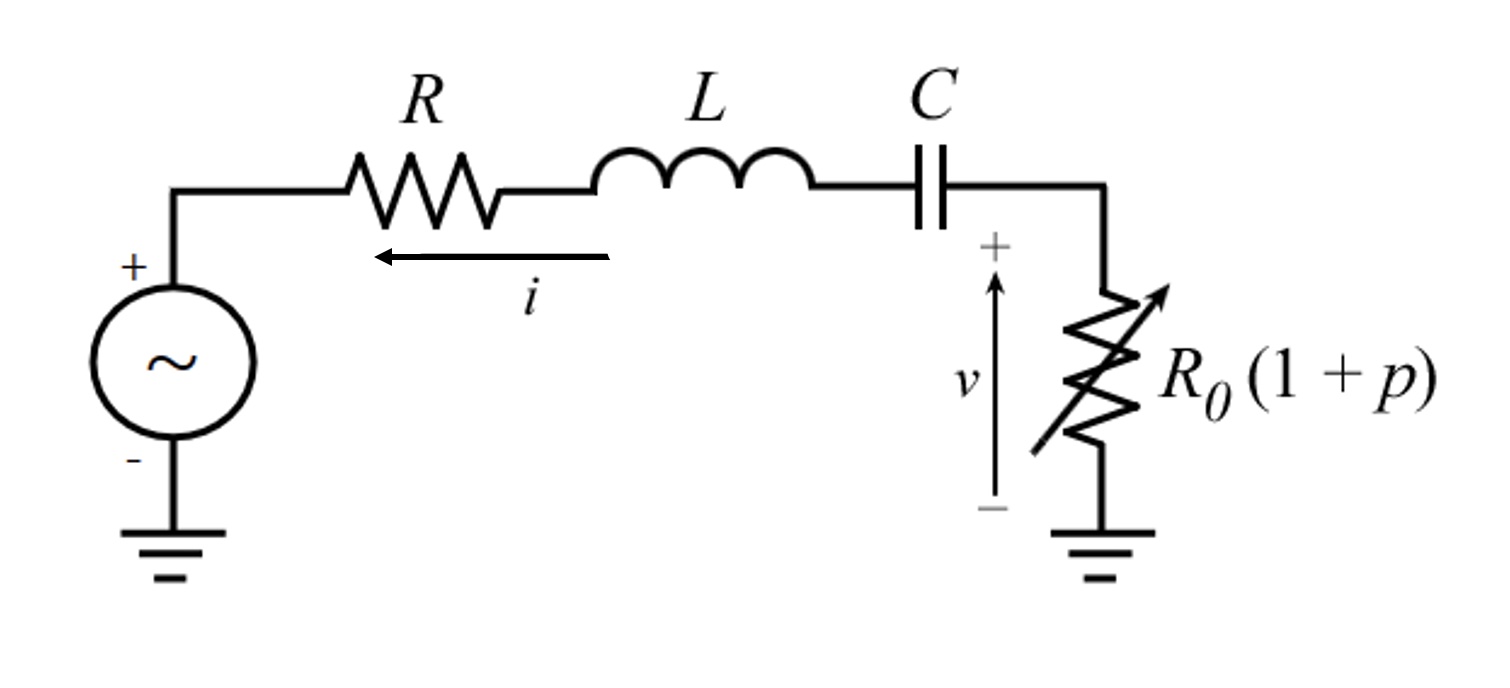}
    \caption{Testbed 1. $R=0.2$ p.u., $X_L =0.25 $ p.u., $X_C = 0.1$ p.u. $R_0 = 10$ p.u. }
    \label{fig:testbed1}
\end{figure}

A testbed of this circuit is built in MATLAB Simscape Specialized Power Systems environment. Fig. \ref{fig:testbed1} shows the testbed along with the GridSweep probing. The parameters of the systems are shown in the figure caption. 
%
 
\begin{figure}[!h]
    \centering
    \begin{subfigure}[b]{3.5in}
    \centering
    \includegraphics[width=0.9\linewidth]{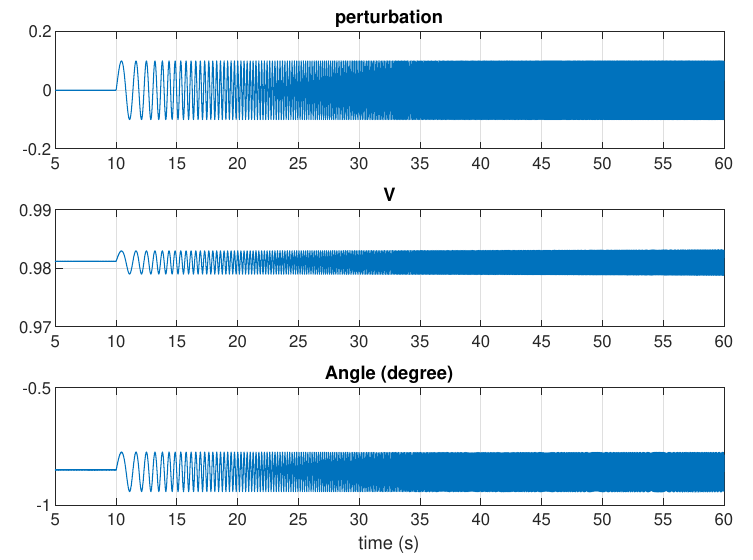}
    \caption{    \label{case1:fig1}}
    \end{subfigure}
 \begin{subfigure}[b]{3.5in}
    \centering
    \includegraphics[width=0.9\linewidth]{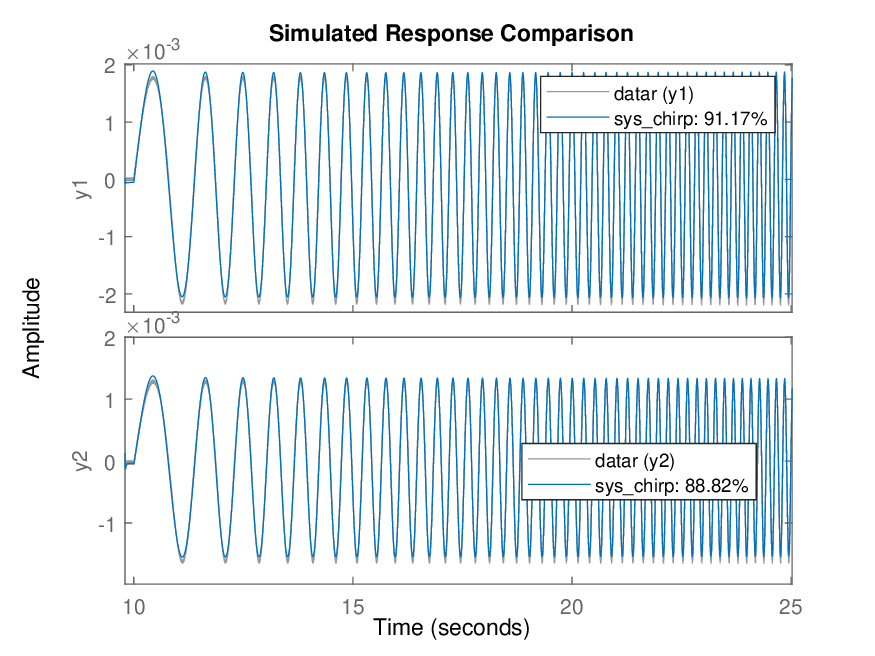}
            \caption{\label{case1:fig2}}
    \end{subfigure}
    \caption{(a) The input chirp signal perturbation and output measurement data of voltage magnitude and angle.  (b) Comparison of the estimated model output and the measured output.}
\end{figure}

Fig. \ref{case1:fig1} shows the chirp signal perturbation employed and the measurements of voltage magnitude and angle. The chirp signal starts at 0.5 Hz and ends at 20 Hz. It is an efficient way to sweep the frequency spectrum, and has the same characteristics of the single sinusoidal perturbation method used by the GridSweep device. This set of input and output data has been used to carry out system identification to find the input/output transfer function. 
 We have used MATLAB System Identification Toolbox's \texttt{tfest} function to conduct identification. The system's order is selected to achieve the best matching between the model's output and the measured output.  For this set of data, the 4th order model yields the best result. Fig. \ref{case1:fig2} shows the comparison results of the measurements and the model's outputs. It can be seen that the matching degrees of 91\% and 89\% are reached.

\begin{figure}[!h]
    \centering
    \begin{subfigure}{0.8\linewidth}
    \includegraphics[width=\linewidth]{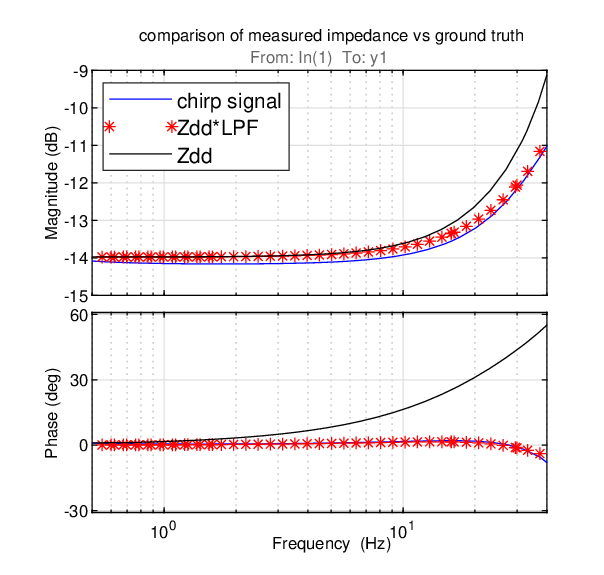}
    \caption{}
    \end{subfigure}
        \begin{subfigure}{0.8\linewidth}
     \includegraphics[width=\linewidth]{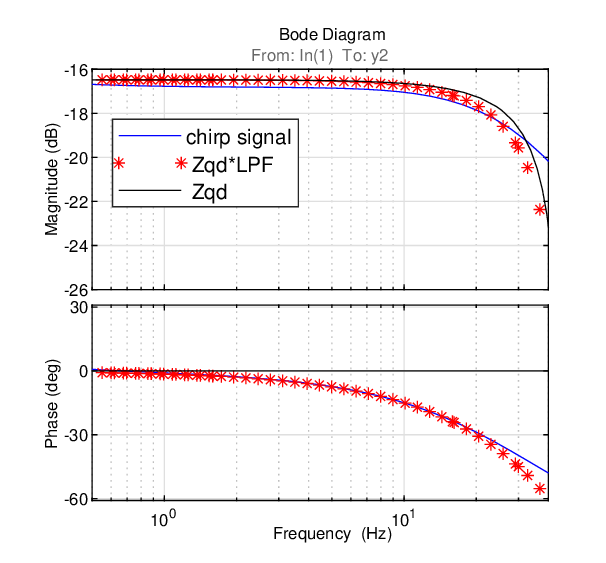}
     \caption{}
         \end{subfigure}
        \caption{Comparison of the impedance components from data and the ground truth: (a) the $dd$ component; (b) the $qd$ component.}
    \label{case1:fig4}
       \vspace{-0.15in}
\end{figure}
The identified model (blue line) is compared with the analytical model $\bf{Z}$'s $dd$ and $qd$ components (black lines). The results are shown in Fig. \ref{case1:fig4}. It can be seen that the identified models lag $Z_{dd}$ and $Z_{qd}$. At 40 Hz, the phase lag is about 60 degrees. Re-examining the phasor extracting process shown in Fig. \ref{fig_dq}, we find that the effect of the LPF has to be considered. 

In Testbed 1, we have used a moving-average filter to process data and filter out the second harmonic component generated by the multiplication of two signals of fundamental frequency. This filter integrates a period of data and delivers the average value. Its transfer function can be expressed as follows: 
\begin{align}
\text{LPF} = \frac{1}{T_w} \left(\frac{1}{s} - \frac{e^{-T_w s}}{s} \right),
\end{align}
where $T_w$ is the time period of the 120-Hz harmonic and $T_w = 1/120$ s. 

Once the effect of the LPF is considered, it can be seen from Fig. \ref{case1:fig4} that the measured impedance components can match $Z_{dd}\cdot {\rm LPF }$ and $Z_{qd} \cdot {\rm LPF}$ very well. 

\emph{Remarks:} This case study demonstrates that the GridSweep approach has the capability to identify the first column of the subsynchronous apparent impedance matrix. 

{
\subsection{Case 2: Measure a circuit with shunt compensation}
In Case 2, we measure a circuit with shunt compensation and further show that the peaks in the measured spectra correspond to the system's eigenvalues or oscillation modes. The RLC circuit has its parameters configured to create an oscillation mode of 69 Hz. In turn, 9-Hz oscillations will be apparent in the voltage magnitude measurements for this system. This 9-Hz mode will be present in the GridSweep's spectrum measurement as a peak at 9 Hz. 
\begin{figure}[!h]
    \centering
   \includegraphics[width=0.9\linewidth]{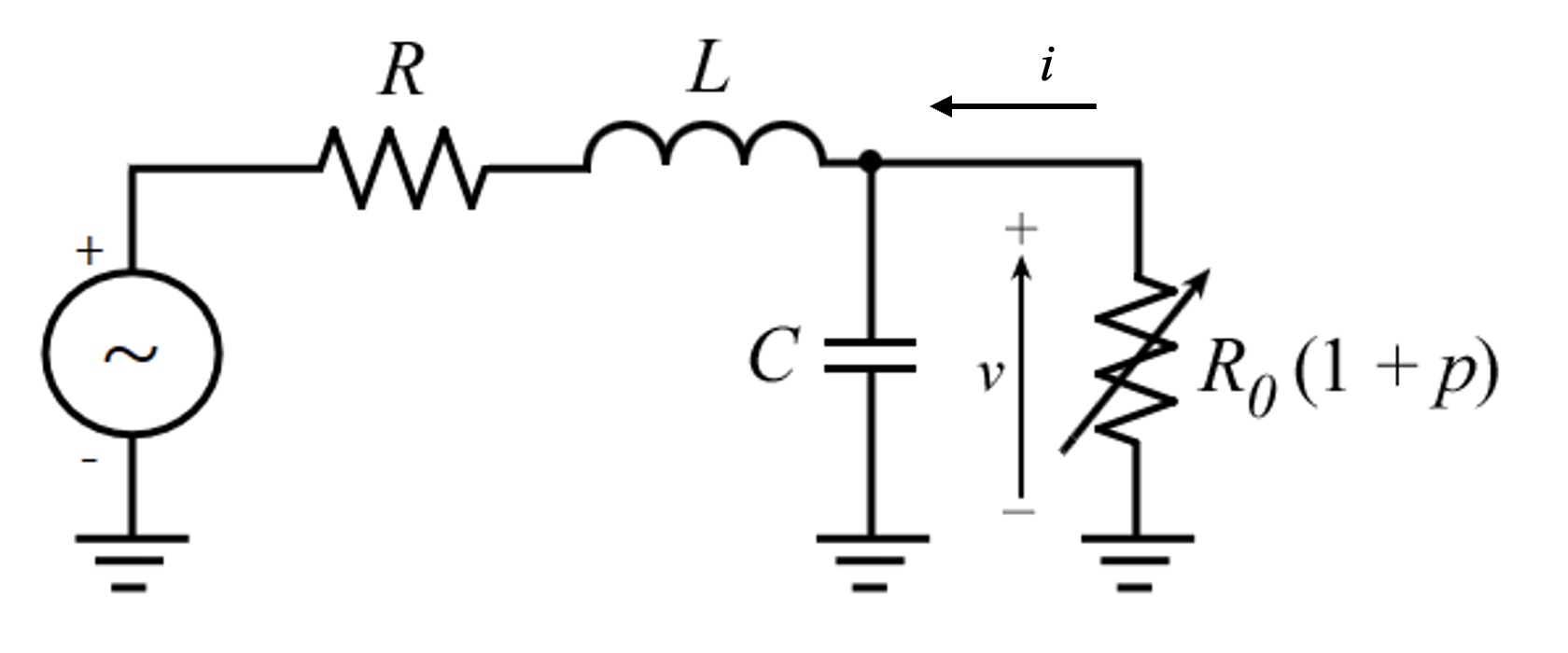}
    \caption{Testbed 2. $R=0.1$ p.u., $X_L =0.75 $ p.u., $B = 1$ p.u. $R_0 = 50$ p.u. }
    \label{fig:testbed2}
\end{figure}

Testbed 2 is shown in Fig. \ref{fig:testbed2}. The relationship between the measured voltage $v$ and the current injection $i$ in the static frame is as follows: 
\begin{align}
Z(s) = \frac{v}{i} = \frac{1}{Cs + \frac{1}{R + Ls}} = \frac{R+Ls}{LC s^2 + RC s + 1}.
\end{align}
For this system, the characteristic polynomial is 
\[s^2 + \frac{R}{Ls} + \frac{1}{LC} = 0.\]

Therefore, the natural oscillation mode has a frequency of 
\begin{align}
\omega_n = \sqrt{\frac{1}{LC}} = \frac{\omega_0}{\sqrt{X B}}.
\end{align}
If $X = 0.75$ and $B = 1$ pu, this mode has 69-Hz as the oscillation frequency. In the $dq$-frame and voltage root-mean square (RMS) or power measurements, the oscillation mode appears as a 9-Hz mode. Additionally, $Z_{dd}$ and $Z_{qd}$ can be found as:
\begin{align}
Z_{dd} = \frac{Z(s+j\omega_0) + Z(s-j\omega_0)}{2}, \notag \\
Z_{qd} = \frac{Z(s+j\omega_0) - Z(s-j\omega_0)}{2j}. 
\end{align}
\begin{figure}[!h]
    \centering
    \begin{subfigure}{0.8\linewidth}
    \includegraphics[width=\linewidth]{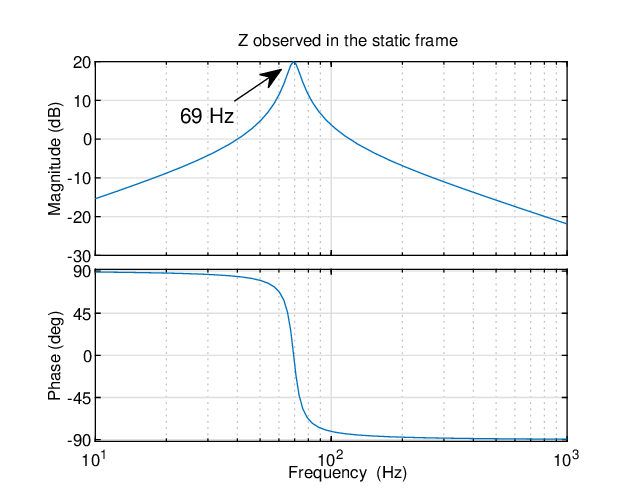}
    \caption{}
    \end{subfigure}
        \begin{subfigure}{0.8\linewidth}
     \includegraphics[width=\linewidth]{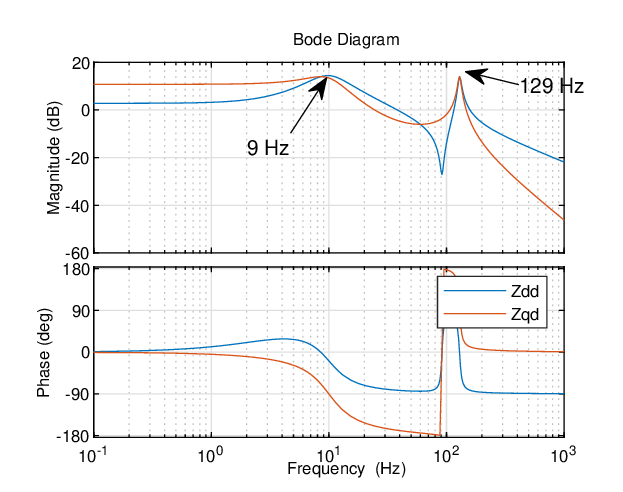}
     \caption{}
         \end{subfigure}
        \caption{The impedance viewed at the probing point for a circuit with shunt compensation. (a) $Z$ in the static frame.  (b) $Z_{dd}$ and $Z_{qd}$.}
    \label{case2:fig1}
    \vspace{-0.15in}
\end{figure}

Fig. \ref{case2:fig1} shows the impedance viewed from the probing point in two frames: the static frame and the $dq$ frame. In the static frame, the impedance shows a peak in its Bode diagram's magnitude at 9 Hz; while in the $dq$ frame, the $dd$ and $qd$ components show peaks at 9 Hz and 129 Hz. Using GridSweep, we expect to see 9 Hz in the measured voltage spectra. To confirm this point, we perturbed the GridSweep's resistance and have a step change of 10\%.

\begin{figure}[!h]
    \centering
    \includegraphics[width=0.8\linewidth]{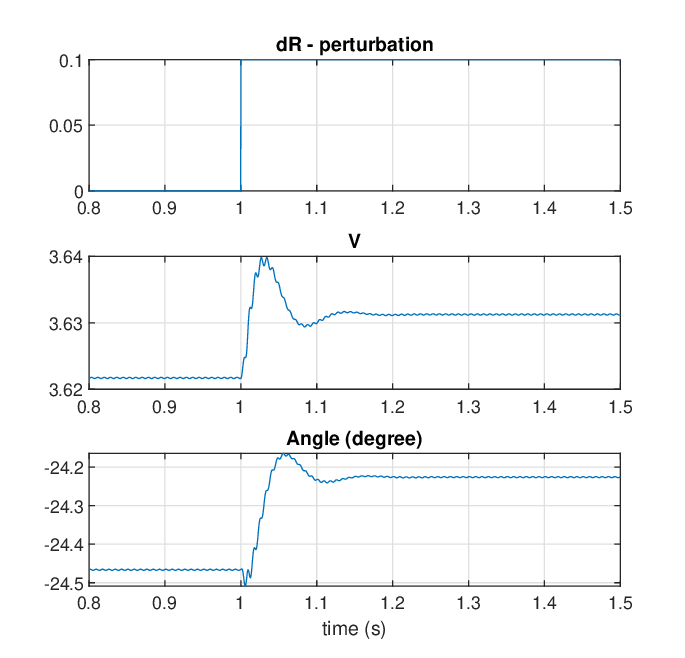}
         \caption{The input and output data to identify the transfer functions of $\hat{v}/p$ and $\theta/p$. }
    \label{case2:fig2}
       \vspace{-0.15in}
\end{figure}
 The nominal GridSweep's resistance is 50 pu. The collected voltage RMS value and the angle measurements are shown in Fig. \ref{case2:fig2}. Based on this set of the data, the input/output models are identified and compared with $Z_{dd}$ and $Z_{qd}$, as shown in Fig. \ref{case2:fig3}. 
\begin{figure}[!h]
\vspace{-0.15in}
    \centering
    \begin{subfigure}{0.8\linewidth}
    \includegraphics[width=\linewidth]{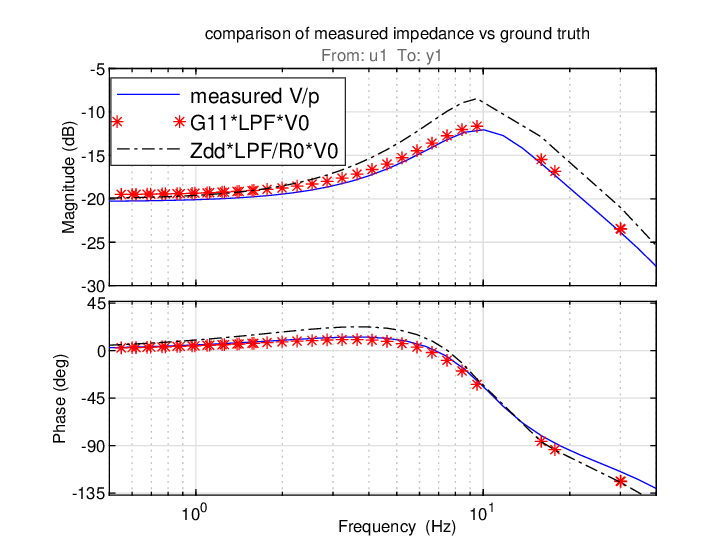}
      \caption{}
    \end{subfigure}
        \begin{subfigure}{0.8\linewidth}
        \includegraphics[width=\linewidth]{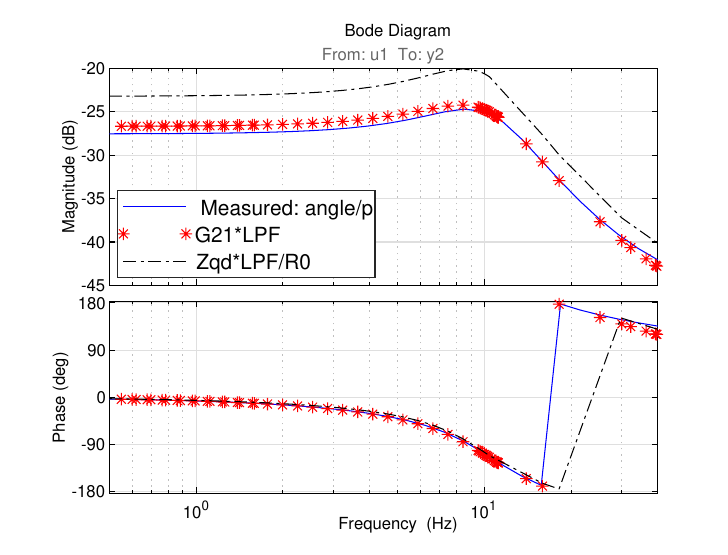}
        \caption{}
        \end{subfigure}
         \caption{The simulated measured impedances vs. the theoretic impedances.  }
    \label{case2:fig3}
    \vspace{-0.15in}
\end{figure}
It can be seen that the measured voltage and angle spectra match very well with the expected results and indeed have peaks at 9 Hz.


 \subsection{Case 3: Multiple voltage measurements}
 \begin{figure}[!h]
    \centering
   \includegraphics[width=1.0\linewidth]{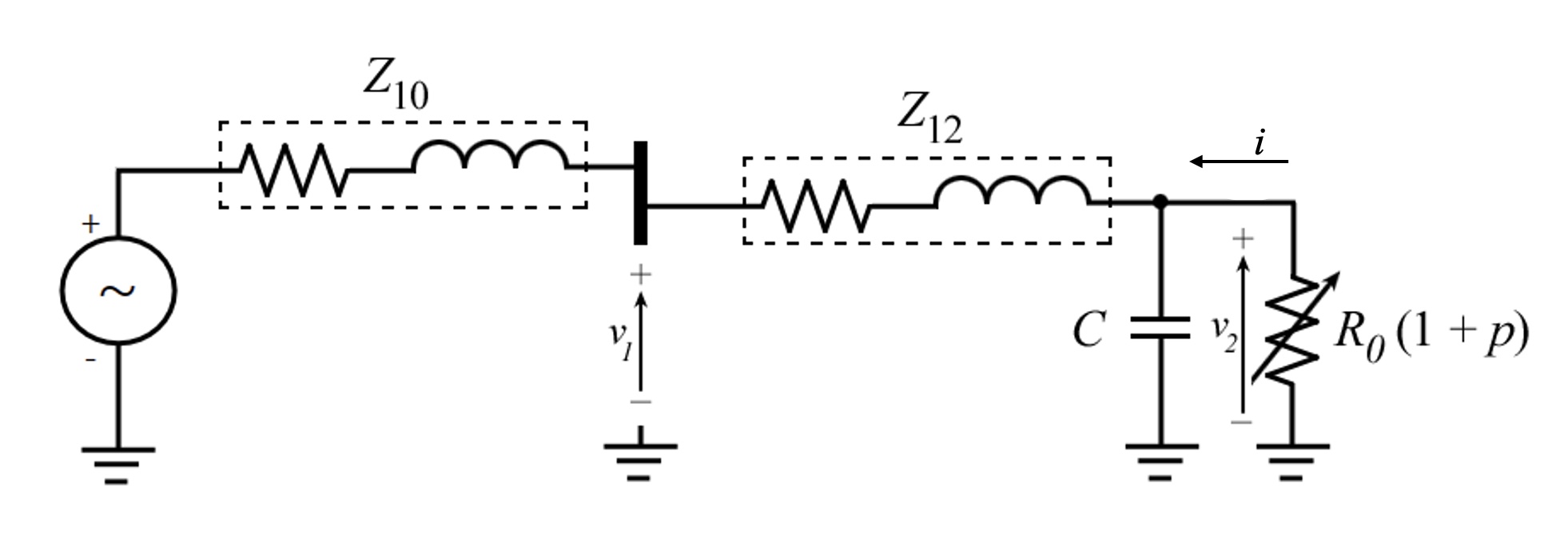}
        \caption{Testbed 3. $Z_{10} = 0.01+0.1/s$, $Z_{12} = 0.09+0.65/s$, $C = 1$ pu. }
    \label{fig:testbed3}
\end{figure}
While the probing is usually placed at the load side, we may take the voltage measurement at a different location. In this subsection, we show the relationship between the two voltages, one taken at the upstream while another taken at the probing location. 

A circuit is used to illustrate the two measurements, as shown in Fig. \ref{fig:testbed3}. The circuit has two buses: Bus 1 and Bus 2, where Bus 2 is the probing location while Bus 1's voltage will also be recorded.

If the impedance between Bus 1 and Bus 2 is $Z_{12}$, and the Bus 1's Th\'evenin equivalent impedance viewed from downstream is $Z_{10}$, we may find the two voltages are related with each other as follows: 
\begin{align}
    \frac{\Delta v_1}{\Delta v_2} = \frac{Z_{10}}{Z_{10}+Z_{12}}. 
\end{align}
If the impedances are in the $dq$-frame and are 2-by-2 matrices, then the relationship between the two voltage vectors are:
\begin{align}
    \Delta {\bf V}_1 &= {\bf Z}_{10}({\bf Z}_{10}+{\bf Z}_{12})^{-1} \Delta {\bf V}_2 \notag \\
   & \approx{\bf Z}_{10}({\bf Z}_{10}+{\bf Z}_{12})^{-1} {\bf Z}_{{\rm apparent},c1}\frac{p}{R_0}{\bf V}_0
   \label{eq15}
\end{align}
where ${\bf Z}_{{\rm apparent},c1}$ notates the first column of the apparent impedance viewed at the probing location: Bus 2. 

Therefore, it can be seen that the voltage measured at the upstream location also contains the information of the apparent impedance.  The measured voltage and angle at Bus 1 and Bus 2 vs. the perturbation at Bus 2 are compared with the results based on the analytical equations in \eqref{eq15}. Fig. \ref{case3:fig} shows the comparison results.
%

\begin{figure}[!h]
    \centering
    \begin{subfigure}{0.80\linewidth}
    \includegraphics[width=\linewidth]{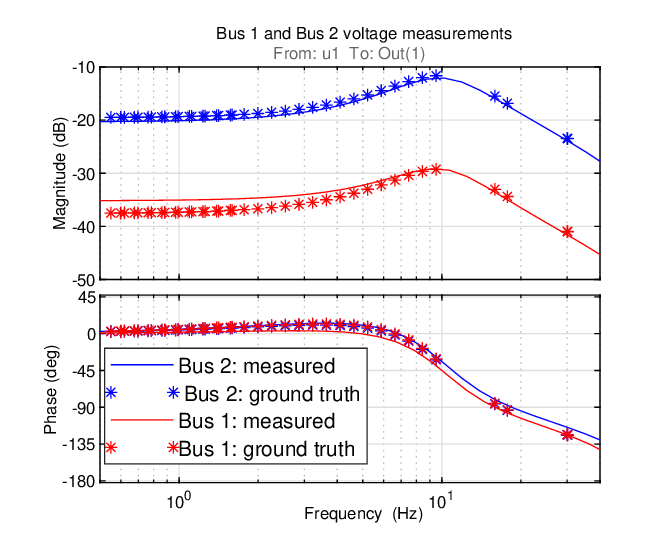}
    \caption{}
    \end{subfigure}
          \begin{subfigure}{0.80\linewidth}
      \includegraphics[width=\linewidth]{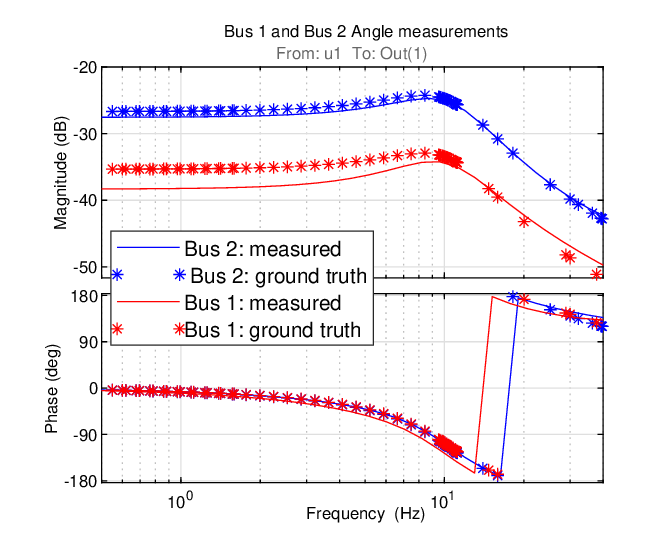}
     \caption{}
         \end{subfigure}
        \caption{Comparison of the impedance components from data and the ground truth: (a) the $dd$ component; (b) the $qd$ component.}
    \label{case3:fig}
\end{figure}

Fig. \ref{case3:fig}(a) shows the voltage spectra at two locations. Fig. \ref{case3:fig}(b) shows the angle spectra at two locations. It can be seen that the measured spectra agree with the ground truth very well. All spectra show a peak at 9 Hz, which implicates that at any location, the measurements can always reflect the closed-loop system poles or the system oscillation modes. 

}
\begin{figure}[!h]
    \centering
    \includegraphics[width=1.0\linewidth]{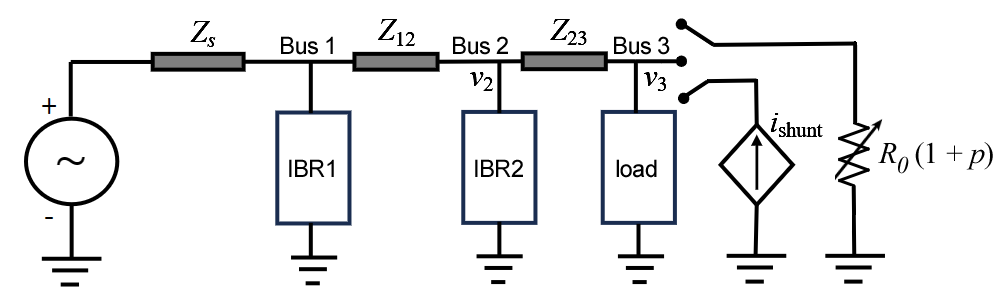}
    \caption{A single-phase distribution feeder with multiple IBRs.}
    \label{fig:testbed4}
\end{figure}
\section{Simulate measurements of distribution systems with IBRs } 
Next, we simulate measuring the apparent subsynchronous impedance of a distribution feeder using multiple simulated IBRs, as shown in Fig. \ref{fig:testbed4}.

\begin{figure}[!h]
    \centering
    \includegraphics[width=1.0\linewidth]{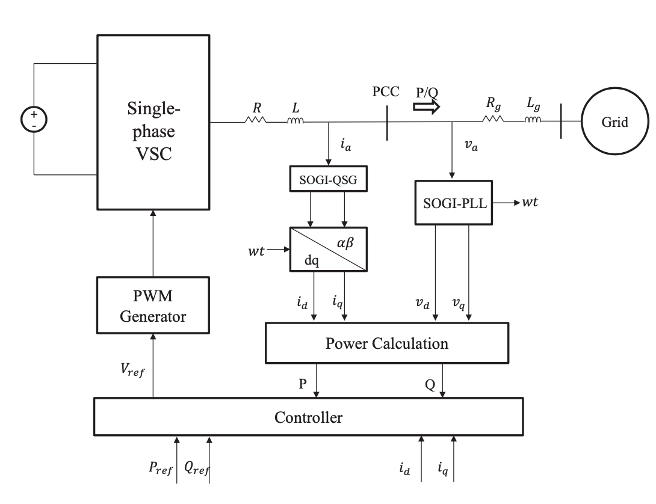}
    \includegraphics[width=1.0\linewidth]{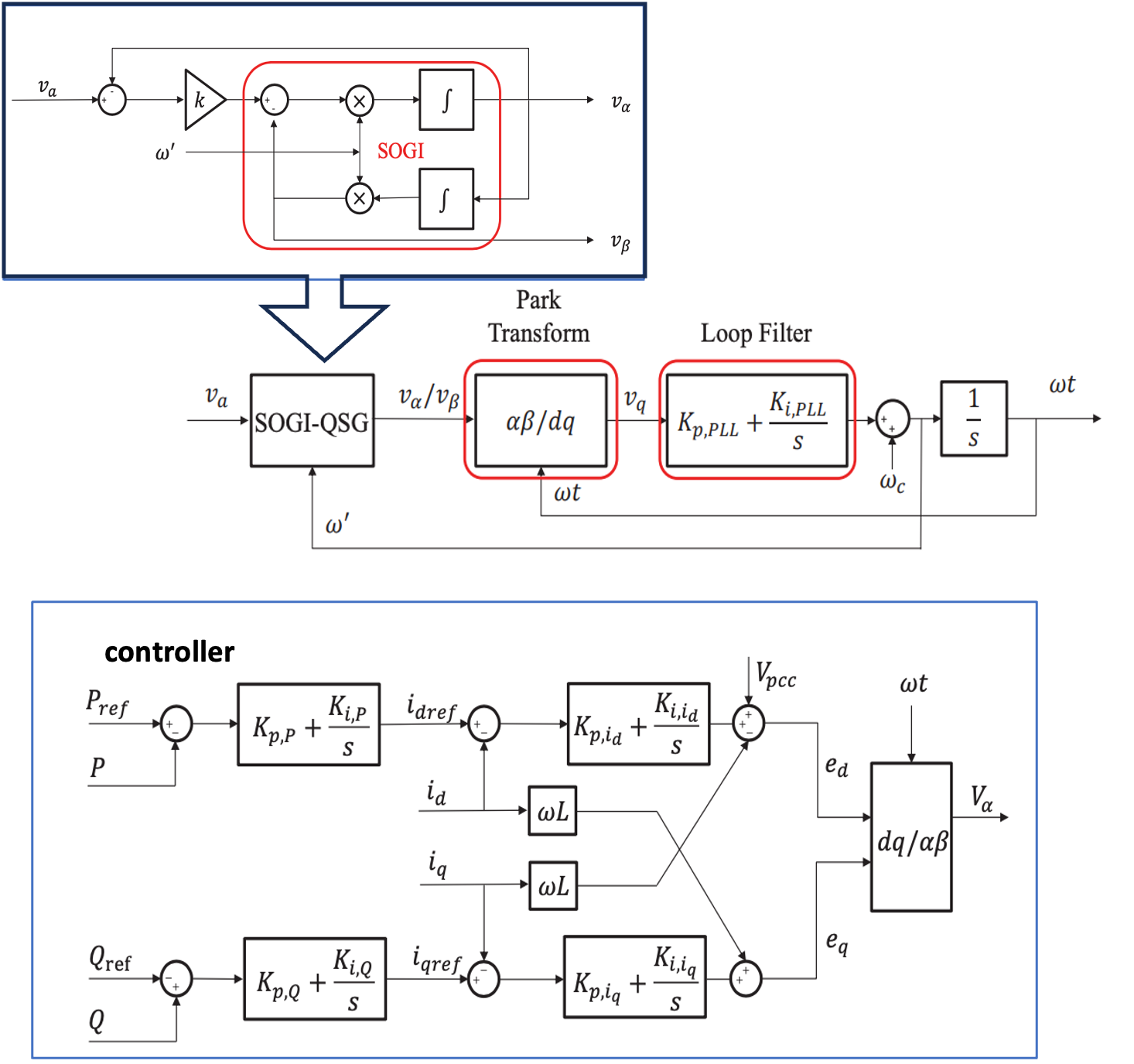}
    \caption{IBR control structure and its circuit topology ($R$ and $L$ are for the choke filter), the SOGI-PLL structure, and the PQ controller structure.}
    \label{fig:distribution1}
\end{figure}

\begin{figure*}[!h]
\centering
\begin{subfigure}{7.2in}
\centering
       \includegraphics[width=7.2in]{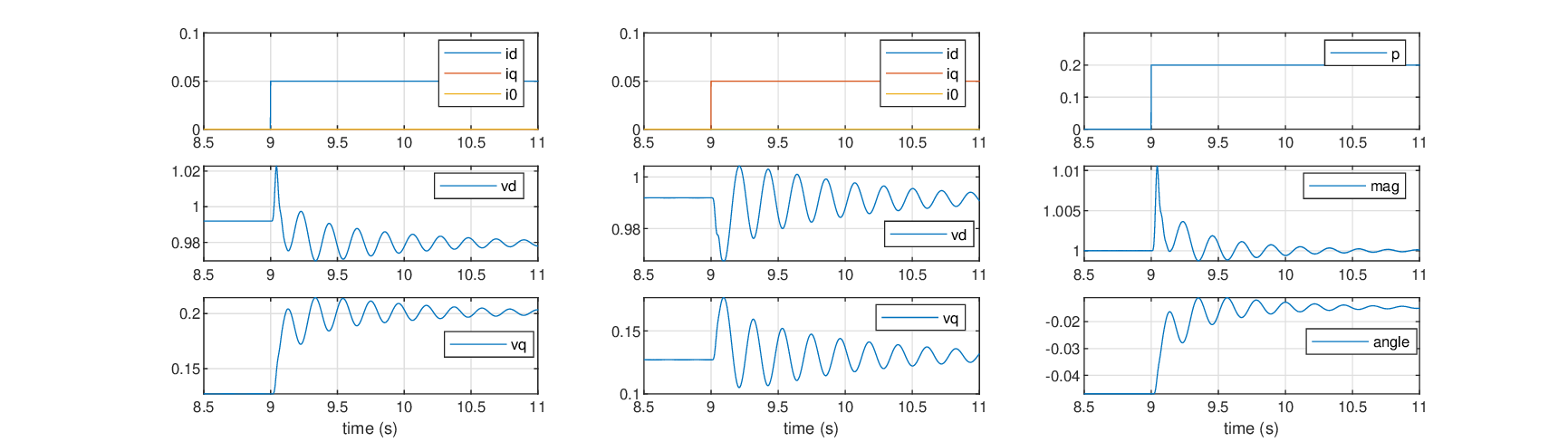}
        \caption{}
    \label{case2:data}
    \end{subfigure}\\
\begin{subfigure}{2.0in}
       \includegraphics[width=2.0in]{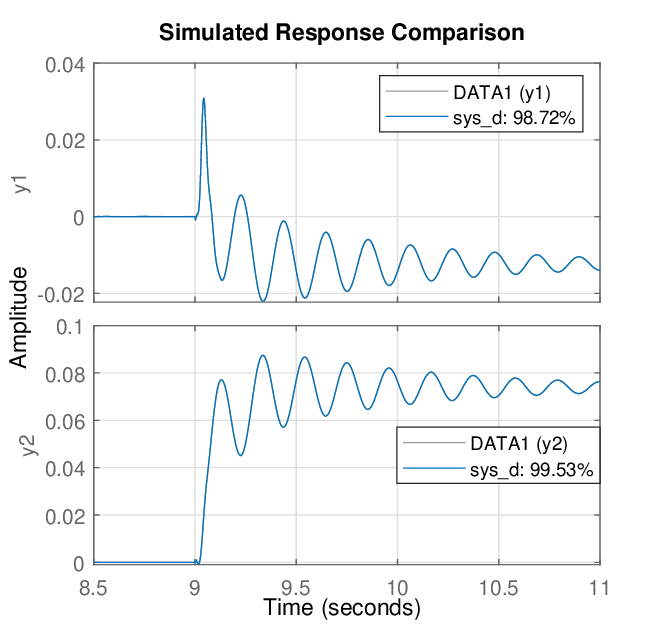}
       \caption{}
       \end{subfigure}
       \begin{subfigure}{2.0in}
              \includegraphics[width=2.0in]{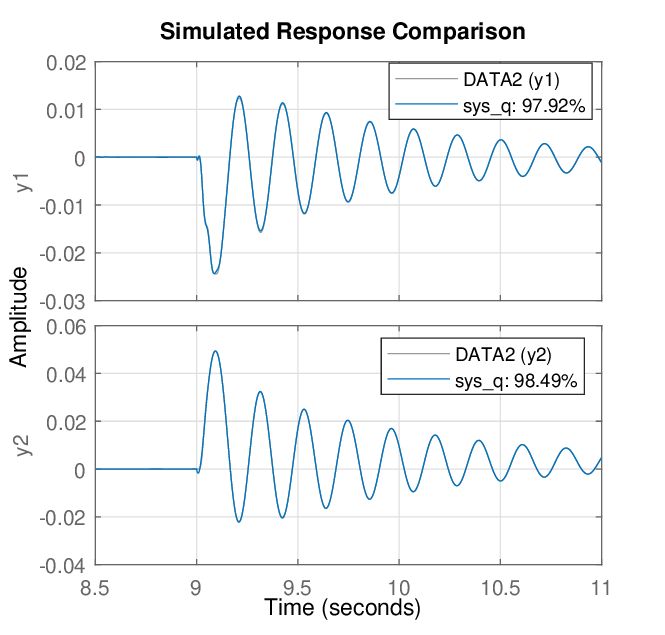}
                    \caption{}
       \end{subfigure}
              \begin{subfigure}{2.0in}
                    \includegraphics[width=2.0in]{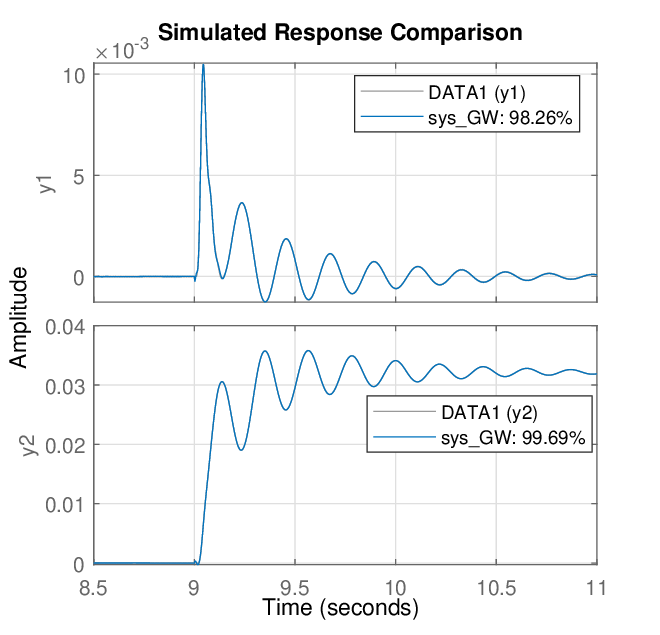}
                          \caption{}
       \end{subfigure}
           \caption{(a) Step response data collected. Columns 1\&2: current injection. Column 3: GridSweep. (b)(c)(d) System identification by use of the input/output data. (b) $d$-axis shunt current injection. (c)$q$-axis shunt current injection. (d) GridSweep. }
    \label{case2:sid1}
\end{figure*}           

Due to the complexity of this configuration, the apparent subsynchronous impedance of the system cannot be found analytically.  So we chose a different method to determine ground truth: compare the results of our simulated GridSweep measurement with the results of a different probing method that relies on a current source to find the apparent subsynchronous impedance (Section \ref{s1}). We simulate with both a weak grid and a strong grid, as shown in Table \ref{table_weak-strong}. As expected, the strong grid does not show significant oscillations, while the weak grid shows oscillations. In Section \ref{s2}, we compare the GridSweep measurements when IBR2's control varies in a weak grid.  In Section \ref{s3}, we examine the GridSweep measurements at different locations of a strong grid. 


\begin{table}[!h]
\caption{Two scenarios of the distribution feeder.}  
\label{table_weak-strong}
\centering
\footnotesize
    \begin{tabular}{lcccc}
    \hline \hline
                     & $Z_s$    & $Z_{12}$ & $Z_{23}$ &  $Z_{\rm load}$  \\ \hline 
   Weak grid & $0.5+j0.25$ & $0.2+j0.2$ & $0.1+j0.1$ & $0.7$   \\
   Strong grid &   $0.1+j0.1$    & $0.2+j0.2$  & $0.1+j0.1$ &  $0.4$  \\
      \hline \hline
    \end{tabular}
  \end{table}
{\subsection{Description of Two Scenarios}

The distribution feeder's line impedances $Z_s$, $Z_{12}$, $Z_{23}$ will be set differently for two scenarios, as shown in Table \ref{table_weak-strong}.

The first scenario in sections \ref{s1} and \ref{s2} demonstrates oscillatory characteristics under weak grid conditions. The apparent impedance shows peaks in its subsynchronous spectra. The peaks imply that the system has poorly damped oscillation modes. 
In contrast, the second scenario in Section \ref{s3} has a relatively stiff feeder. Simulated measurements show the lack of peaks in the subsynchronous spectra, implying low risk of oscillation.  

In our simulation the single-phase distribution feeder's nominal RMS voltage is set to 190.52 V and the power rating is set as 4-kW. This is also the rating of the two IBRs. The single-phase IBR’s dc side is directly connected to a 400-V dc voltage source. Fig. \ref{fig:distribution1} shows the overall control structure of a grid-following converter and the circuit topology. The control has two main blocks: the synchronizing unit and the controller that regulate real power and voltage (or reactive power).

The detailed synchronizing unit and the PQ controller are also shown in Fig. \ref{fig:distribution1}. The synchronizing unit is a second-order generalized integral-phase-locked-loop (SOGI-PLL). This type of PLL is suitable for single-phase synchronization as it uses a single voltage input, generates a copy of the input and its Hilbert transform. Both become the input and are converted to the $dq$-frame variables by use of the PLL’s output angle. Enforcing the $q$-axis component to zero aligns the voltage space vector with the $d$-axis of the PLL frame. Therefore, the PLL frame tracks the voltage space vector and the PLL’s output angle aligns with the voltage phase angle at steady state.

Table \ref{table1} lists the IBR control and circuit parameters. 
\begin{table}[!t]
\caption{Circuit and control parameters for IBRs.}  
\label{table1}
\centering
\footnotesize{
    \begin{tabular}{l|l|l}
    \hline \hline
    Description & Parameters & Values \\\hline 
    Power Base & $S_b$ & 4 kW \\    
   Converter Voltage Base & $V_b$ & 190 V \\   
    Nominal Frequency & $\omega_0$ & 2$\pi$60 rad/s \\    
      Grid Voltage & $V_g$ & 1 pu \\   
    Choke Filter            &  $R$, $L$ &  $0.005$, $0.15$ pu\\
     Inner Loop Control & $k_{ip}$, $k_{ii}$ & {0.3, 5} pu\\     
    $P$ control & $k_{Pp}$, $k_{Pi}$ &{0.4, 40}  pu\\
     $V$ or $Q$ control & $k_{Vp}$, $k_{Vi}$ &{0.4, 40} pu \\
      PLL & $k_{p\text{PLL}}$, $k_{i\text{PLL}}$ &{100, 1400} pu \\
    \hline \hline
    \end{tabular}
   }
\end{table}

}
\subsection{Comparison of GridSweep Measurements vs. Current Injection-based Measurements}
\label{s1}
This system is complicated, so it is difficult to find an analytical form of the apparent impedance. Instead, as a reference, we use current injection perturbation and voltage measurements to identify the apparent impedance, similar to \cite{rygg2017apparent}. This measured impedance is treated as the benchmark to provide a comparison against the impedance found from GridSweep. 

\begin{figure}[!h]
\centering
       \includegraphics[width=3.0in]{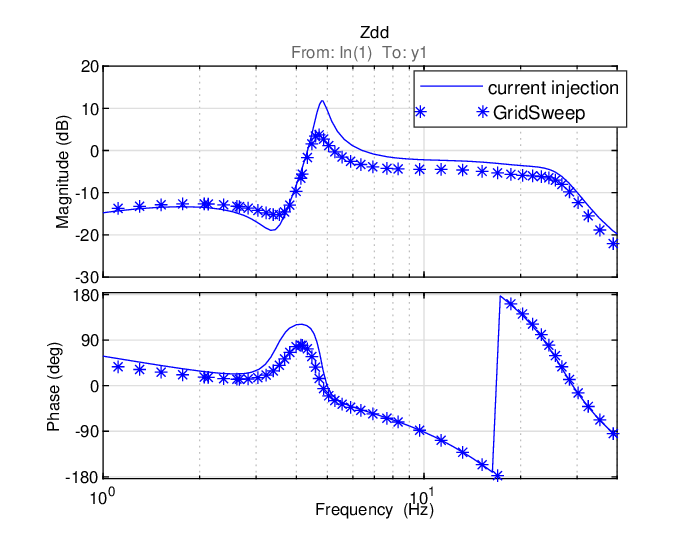}
       \includegraphics[width=3.0in]{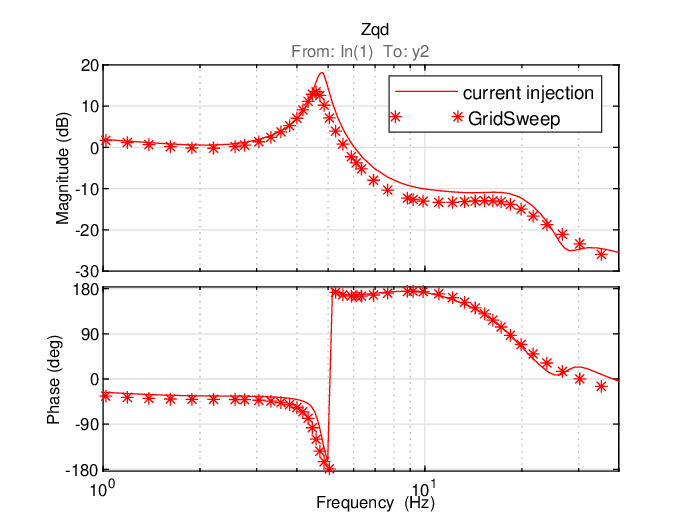}
           \caption{Comparison of the impedance measurements from the two methods: current injection and GridSweep.  }
    \label{case2:sid2}
\end{figure}  

The testbed is shown in Fig. \ref{fig:testbed4}, along with the current injection perturbation as well as the GridSweep perturbation. The current injection perturbation is conducted in two steps. First, the $d$-axis current is perturbed using a step change.  The $dq$-axis voltage measurements are recorded. Then, the $q$-axis current is perturbed by a step change and the $dq$-axis voltage measurements are recorded. The two sets of the data are used for model identification, as shown in Fig. \ref{case2:sid1}. The $dq$ frame is a synchronous rotating frame aligned with the grid voltage. Therefore, the measured impedance is based on the grid $dq$ frame. On the other hand, the impedance that GridSweep can observe is based on the probing site voltage. This $dq$ frame and the grid $dq$ frame may have an angle $\delta$, and $\delta$ influences the two impedances, as follows: 
\begin{align}
{\bf Z}_{GW} = T {\bf Z}_{\rm grid} T^{-1}, 
\end{align}
where ${\bf Z}_{GW}$ is the impedance that can be detected by GridSweep while ${\bf Z}_{\rm grid}$ is the impedance based on the grid $dq$ frame. $T$ is the transformation matrix and the above relationship has been derived in \cite{fan2020modular}.
\begin{align}
T = \begin{bmatrix}   \cos\delta & -\sin\delta \\ \sin\delta & \cos\delta  \end{bmatrix}
\end{align}

%
Fig. \ref{case2:sid2} shows the comparison of the $dd$ and $qd$ components of the impedance based on the two methods: current injection and GridSweep. Both impedances are based on the $dq$ frame aligned with the voltage at the local probing bus.  It can be seen that both methods lead to comparable results. We conclude that the GridSweep method is capable of identifying the subsynchronous apparent impedance. 

{
\subsection{Comparison of GridSweep Measurements for Different IBR Controls}
\label{s2}
We further compare the GridSweep simulated measurements for the same distribution feeder, except that IBR2's $q$-axis outer control has been changed from voltage control to reactive power control. In the first case, the distribution feeder has a damped oscillation mode at 4.5 Hz; while in the second case, the distribution feeder has a poorly damped oscillation mode at 6 Hz. Fig. \ref{twomodes}(a) show the simulated GridSweep perturbation and output data. The measured spectra based on the sets of the data are shown in Fig. \ref{twomodes}(b). It can be clearly seen that the spectra show a sharp peak at 6 Hz implicating poorly damped oscillations.  
\begin{figure}[!h]
\centering
      \begin{subfigure}{3.0in}
       \includegraphics[width=3.0in]{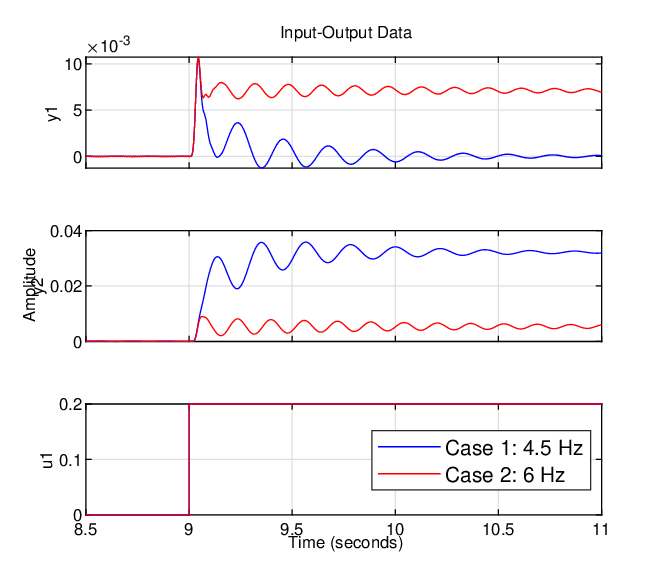}
              \caption{}
       \end{subfigure}
       \begin{subfigure}{3.0in}
       \includegraphics[width=3.0in]{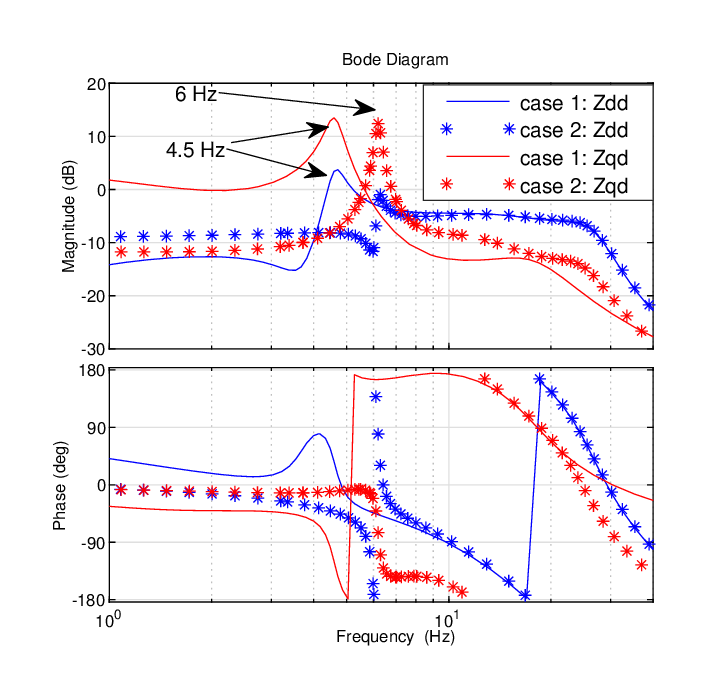}
                     \caption{}
       \end{subfigure}
           \caption{Comparison of the GridSweep measurements for damped 4.5-Hz oscillations vs marginally damped 6-Hz oscillations. (a) The perturbation and the output data. (b) The resulting spectra.  }
    \label{twomodes}
\end{figure}  

The time-domain data in Fig. \ref{twomodes}(a) show that the voltage magnitude has a larger steady-state increase for case 2; while the angle has a greater steady-state increase for case 1. This observation matches the $Z_{dd}$ and $Z_{qd}$'s characteristics for the two cases in the low-frequency region. Additionally, the time-domain data show 4.5-Hz damped oscillation in case 1 and 6-Hz poorly damped oscillations in case 2. This observation also matches the peaks in $Z_{dd}$ and $Z_{qd}$ for the two cases. For case 1, the peak appears at 4.5 Hz, while for case 2, the peak appears at 6 Hz. The 6-Hz peak is much sharper compared to the 4.5-Hz peak, implicating a much more poorly damped oscillation mode.   

We conclude, again, that the subsynchronous spectra measurement of GridSweep method can capture the subsynchronous oscillation modes in an IBR-intense feeder. 

\subsection{Comparison of voltage measurements at different locations}
\label{s3}
In this comparison, the distribution feeder has its impedances configured to represent a strong grid condition. In this scenario, the system experiences minimal oscillations. The voltage measurements at two locations (Bus 2 and Bus 3) are recorded for analysis. Note that Bus 3 is the probing location, and the measured voltage and angle spectra directly indicate the apparent impedance viewed at Bus 3. Bus 2's measurements reflect the apparent impedance viewed at Bus 3 multiplied by a ratio of Bus 2's upstream impedance ($Z_2$) vs $Z_2+Z_{23}$:\
\begin{align}
\text{ratio} = \frac{Z_2}{Z_2 +Z_{23}}. 
\end{align}
$Z_2$ is the impedance viewed from the Bus 2's terminal of Line 2-3. $Z_2$ can be expressed as follows:
\begin{align}
Z_2 &= (Z_{12}+ Z_1)//Z_{\rm IBR2}, \\
Z_1 & = Z_s // Z_{\rm IBR1}.
\end{align}

This ratio is always less than 1. Therefore, the magnitudes of the spectra at Bus 2 are less compared to those at Bus 3. 
\begin{figure}[!h]
\centering
       \begin{subfigure}{2.8in}
       \includegraphics[width=2.8in]{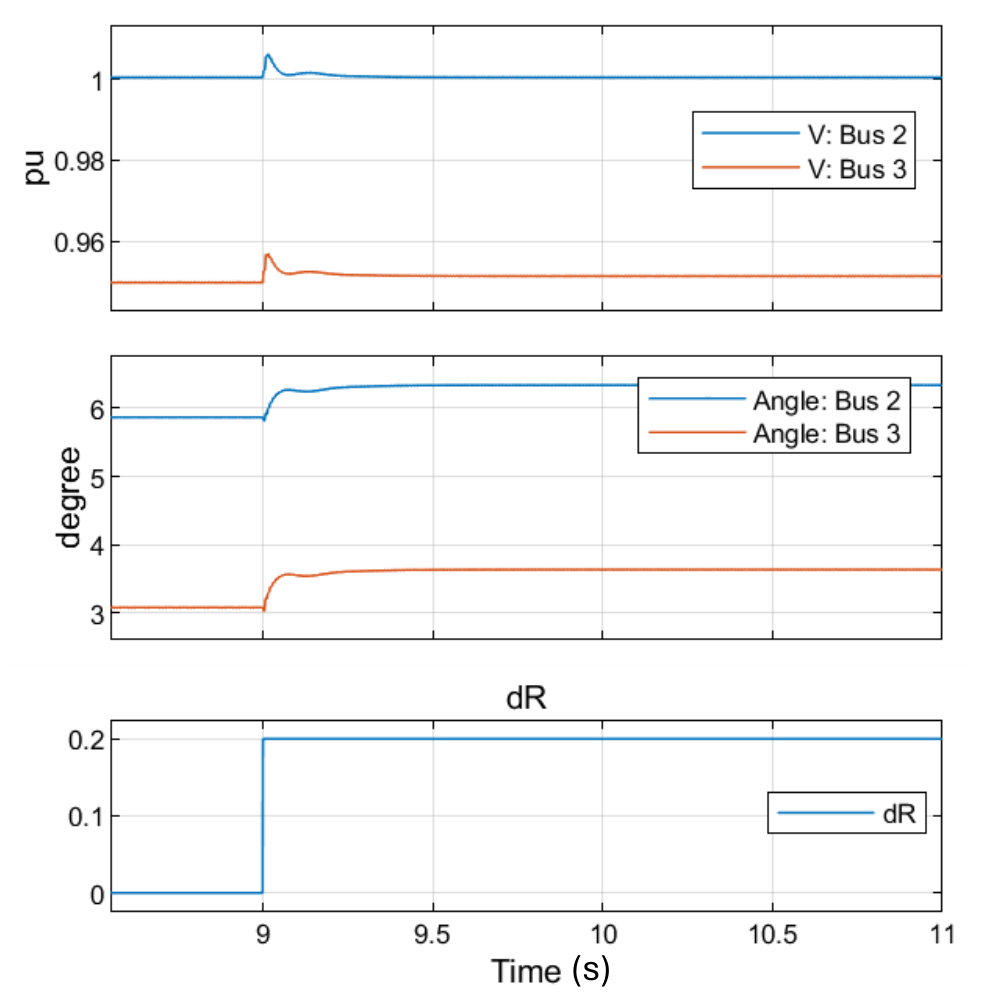}
                   \caption{}
       \end{subfigure}
       \begin{subfigure}{3.0in}
       \includegraphics[width=3.0in]{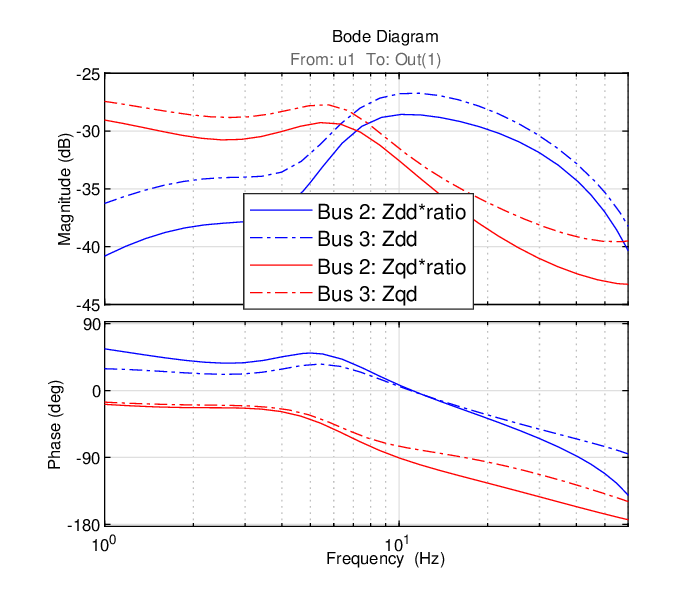}
                         \caption{}
       \end{subfigure}
           \caption{Comparison of the voltage measurements at Bus 2 and Bus 3. (a) Time-domain data. (b) Identified impedance components.  }
    \label{twoV}
\end{figure}  

This point has been confirmed by Fig. \ref{twoV}. Fig. \ref{twoV}(a) shows the perturbed responses collected at Bus 2 and Bus 3, while Fig. \ref{twoV}(b) shows the measured impedance components. 
It is important to note that the overall shape for the frequency response is visible from both measurements, suggesting that if there were an oscillatory mode it would be visible from both locations.

%
\section{Conclusion}
{This paper examines GridSweep's probing and measuring approach for determining the subsynchronous }apparent impedance of a distribution grid. Using simple cases where the apparent impedance is can be explicitly calculated, we demonstrate that by extracting the continuous point-on-wave  voltage measurement's $dq$-frame phasor and identifying the relationship between the voltage phasor and the perturbation in current, the GridSweep approach provides an accurate subsynchronous apparent impedance spectra. In the case of two simple RLC circuits, the apparent impedance spectra from the GridSweep simulation is benchmarked with the calculated ground truth and shows excellent match. The peaks in the subsynchronous impedance spectra match the system's subsynchronous oscillation modes. 
In the case of a more complicated feeder with inverter-based resources, the GridSweep simulation is benchmarked against a conventional current injection method, and shows that both produce the same subsynchronous impedance spectra. Additionally, we demonstrate that the voltage measurements taken at a location different from the probing bus will lead to the subsynchronous apparent impedance multiplied by a ratio. 
Because the GridSweep approach provides a method of measuring the subsynchronous impedance of a distribution feeder using only 120-volt outlet connections, we conclude that this measurement method may well be useful for evaluating risk of oscillation in distributed IBR deployments.
}

\bibliographystyle{IEEEtran} 
\bibliography{ref}

\end{document}